\documentclass[twocolumn,showpacs,preprintnumbers,amsmath,amssymb,prl]{revtex4-1}

\usepackage{dcolumn}% Align table columns on decimal point
\usepackage{upgreek} %\upmu
\usepackage{txfonts} %\muup
\usepackage{ifpdf}
\ifpdf
  \usepackage[pdftex]{graphicx}
 \else
   \usepackage[dvips]{graphicx}

\fi % Used for figures
\usepackage{enumerate}
\usepackage{hyperref}

\begin{document}

\title{Strong polaritonic interaction between flux-flow and phonon resonances in Bi$_2$Sr$_2$CaCu$_2$O$_{8+x}$ intrinsic Josephson junctions:
Angular dependence and the alignment procedure.}

\author{H. Motzkau}
\author{S. O. Katterwe}
\author{A. Rydh}
\author{V. M. Krasnov}
\email{Vladimir.Krasnov@fysik.su.se}
\affiliation{Department of Physics, Stockholm University,
AlbaNova University Center, SE-106 91 Stockholm, Sweden}

\date{\today}

\begin{abstract}
Bi$_{2}$Sr$_{2}$CaCu$_{2}$O$_{8+x}$ single crystals represent
natural stacks of atomic scale intrinsic Josephson junctions,
formed between metallic CuO$_2$-Ca-CuO$_2$ and ionic insulating
SrO-2BiO-SrO layers. Electrostriction effect in the insulating
layers leads to excitation of $c$-axis phonons by the ac-Josephson
effect. Here we study experimentally the interplay between and
velocity matching (Eck) electromagnetic resonances in the
flux-flow state of small mesa structures with $c$-axis optical
phonons. A very strong interaction is reported, which leads to
formation of phonon-polaritons with infrared and Raman-active
transverse optical phonons. A special focus in this work is made
on analysis of the angular dependence of the resonances. We
describe an accurate sample alignment procedure that prevents
intrusion of Abrikosov vortices in fields up to 17\,Tesla, which is
essential for achieving high-quality resonances at record high
frequencies up to 13\,THz.

\end{abstract}


\maketitle

Intrinsic Josephson junctions (IJJs) \cite{Kleiner94} in high
temperature superconductors are considered as possible candidates
for realization of high power THz-frequency radiation sources
\cite{Ozyuzer,Wang,Cascade,Klemm2010,Hu,FiskeTheory}. IJJs can
provide several advantages: (i) A large number of IJJs can
easily be integrated. A strong superradiant emission can take place
if all IJJs are phase-locked in the symmetric in-phase mode. (ii)
Due to the large energy gap in Bi-2212 $\Delta \sim 30-40\,\mathrm{meV}$
\cite{SecondOrder}, IJJs can comfortably operate in the frequency
range up to $\sim 20\,\mathrm{THz}$. (iii) Stacking of IJJs may lead to
cascade amplification of non-equilibrium population, needed for
achieving population inversion and lasing \cite{Cascade}. (iv) A
large variety of emission mechanisms are available. The
ac-Josephson effect facilitates electromagnetic wave emission via
fluxon-induced self-oscillations at geometrical resonances
\cite{Hu,SelfOscill}, or via the Josephson flux-flow emission
\cite{FiskeTheory,Superluminal}. Besides, coherent phonon emission
can take place either without the ac-Josephson effect via
recombination of non-equilibrium quasiparticles \cite{Cascade}, or
via the ac-Josephson effect due to the electrostriction phenomenon
\cite{Polariton}.

In this work we investigate the mutual interaction between phonons
and electromagnetic waves in the flux-flow state of small Bi-2212
mesa structures. The ac-Josephson effect leads to excitation of
$c$-axis optical phonons in Bi-2212 single crystal by means of
electrostriction: Ions in the polar insulating SrO-2BiO-SrO
layers, where the oscillating electric field is concentrated, are
shaken at the Josephson frequency. Both Raman and infrared (IR) -
active $c$-axis phonons can be excited
\cite{Schlen,Ponomarev,Polariton}. The symmetry of the phonon is
directly related to the symmetry of the electromagnetic mode:
Raman phonons are excited by the out-of-phase mode, while
infrared-active phonons are excited by the in-phase mode. The strongest
interaction between electromagnetic waves and phonons occurs at
the transverse optical phonon frequency. This leads to formation
of phonon-polaritons with very slow group velocities. We argue
that resonances with IR-active phonons can stabilize the in-phase
electromagnetic mode in IJJs, needed for achieving coherent
superradiant flux-flow emission from Bi-2212 mesas. The appearance
of phonon-polaritons also demonstrates the existence of an unscreened
polar response, which may lead to the strong electron-phonon
coupling in cuprates.

\begin{figure*}[t]
\begin{center}
   \includegraphics[width=0.9\textwidth]{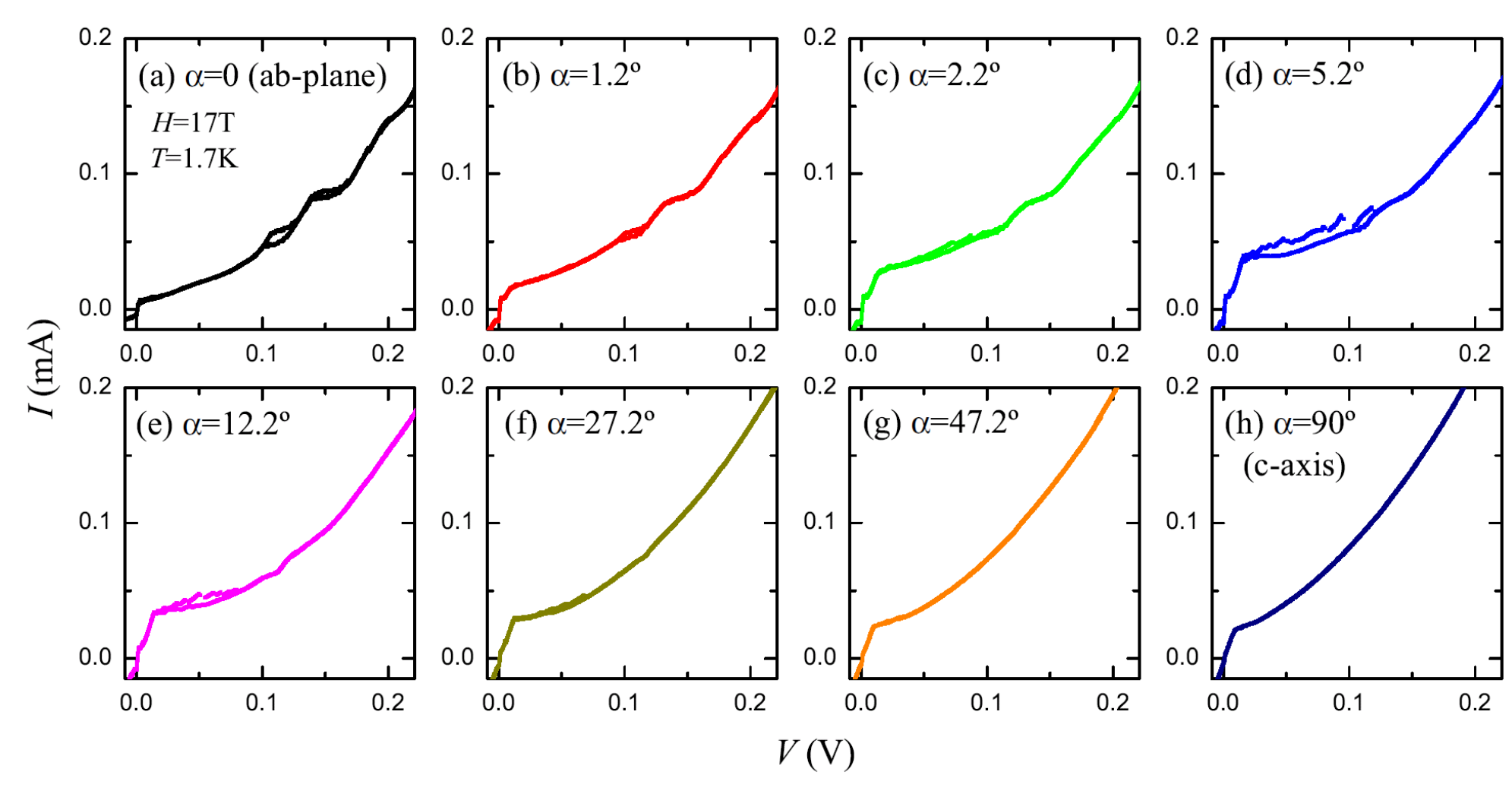}
\end{center}
\caption{\label{fig1} (Color online). Angular dependence of
current-voltage characteristic for a Bi(Pb)-2212 mesa P22/m5b at
$T=1.7\,\mathrm{K}$ and $H=17\,\mathrm{T}$. }
\end{figure*}

We study small mesa structures, made at the surface of freshly
cleaved Bi-2212 single crystals. Mesas are made by
photolithography, argon milling and focused ion beam trimming down
to sub-micron sizes. Two types of crystals are studied: slightly
underdoped pure Bi-2212 crystals with $T_\mathrm{c} \simeq 82\,\mathrm{K}$ and
slightly overdoped lead-doped
Bi$_{2-y}$Pb$_y$Sr$_{2}$CaCu$_{2}$O$_{8+x}$ (Bi(Pb)-2212) crystals
with $T_\mathrm{c} \simeq 90\,\mathrm{K}$. The most noticeable difference between
those crystals is in the critical current density, which is an
order of magnitude larger for Bi(Pb)-2212 ($J_\mathrm{c}\sim 10^4\,\mathrm{A/cm^2}$).
Since the amplitudes of the phonon resonance, reported
below, are proportional to $J_\mathrm{c}^2$ \cite{Heim}, they are much more
pronounced in Bi(Pb)-2212 crystals \cite{Polariton}. In order to
avoid repetitions we address the reader to our previous works, in
which we studied properties of similar mesas in intense magnetic
fields, including: the Fraunhofer modulation of the Josephson
current \cite {Katterwe}, velocity matching (Eck), geometrical
(Fiske) \cite{Superluminal} and phonon-polariton \cite{Polariton}
resonances in the flux-flow state, their temperature dependence
\cite{TempDep} and the intrinsic tunneling magnetoresistance of
IJJs \cite{MR}. Below we will focus on the angular dependence and
describe the procedure for accurate sample alignment, which is
a prerequisite for observation of high quality resonances.

\begin{figure}[b]
\begin{center}
   \includegraphics[width=0.4\textwidth]{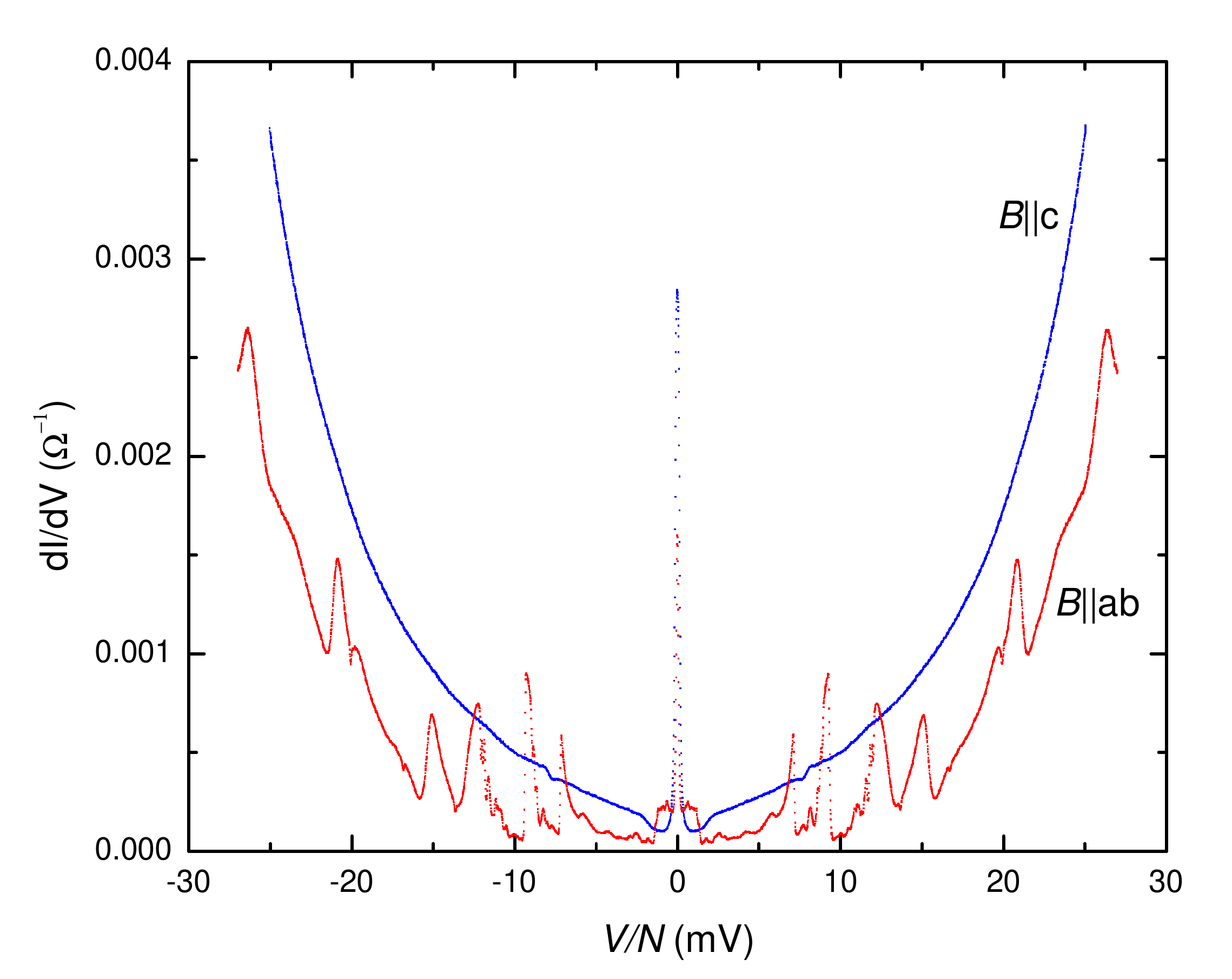}
\end{center}
\caption{\label{fig2} (Color online). Tunneling conductance versus
voltage per junction for a Bi(Pb)-2212 mesa P22/m4a at $T=20\,\mathrm{K}$
and $H=16\,\mathrm{T}$ in the $ab$-plane and in the $c$-axis direction.}
\end{figure}

Figure \ref{fig1} represents the angular dependence of the current-voltage
characteristic of the Bi(Pb)-2212 mesa P22/m5b with the area
$\sim 1.3 \times 0.9\,\mathrm{\mu m^2}$, containing $N \simeq 11$ IJJs. For
the field parallel to the $ab$-plane, (a) $\alpha=0$, the
quasiparticle (QP) branches are completely suppressed and a
pronounced flux-flow step appears at $V \simeq 0.1\,\mathrm{V}$. It
corresponds to the Eck resonance, when the speed of propagating
Josephson vortices (fluxons) is equal to the speed of the slowest
out-of-phase electromagnetic wave
\cite{Superluminal,Polariton,TempDep}. Strong hysteretic
resonances above the Eck step at higher voltages, $0.1\,\mathrm{V}$
$\lesssim V \lesssim 0.2\,\mathrm{V}$, are due to excitation of $c$-axis
optical phonons \cite{Schlen,Heim,Preis,Ponomarev,Polariton}.
Already at a slight misalignment of the field by 1-2 degrees (b,
c) the resonances are strongly dampened due to penetration of
Abrikosov vortices. The amount of Abrikosov vortices increases
with further increase of the angle (d), they effectively pin
Josephson fluxons, which leads to a significant increase of the
critical current, suppression of the flux-flow and a reappearance of QP
branches (d, e). With further increase of the angle the amount of
Abrikosov vortices becomes so large that their normal cores start
to suppress the superconducting order parameter. This results in
the pronounced negative magnetoresistance \cite{MR} and the
suppression of the critical current (f-h).

Figure \ref{fig2} represents differential conductance curves $\mathrm{d}I/\mathrm{d}V$ versus
voltage per junction $V/N$ for another Bi(Pb)-2212 mesa (P22/m4a
$1.2 \times 2.0 ~\mu m^2$, $N \simeq 55$) for field along the
$ab$-plane and in the $c$-axis direction. Phonon resonances are
seen as peaks in $\mathrm{d}I/\mathrm{d}V$ for a field along the $ab$-plane. They
completely disappear for field in the $c$-axis direction, clearly
showing the importance of sample alignment.

\begin{table*}
   \begin{ruledtabular}
       \begin{tabular}{ccccccccccccccc}
       $n$&                       1&     2&  3&       4&    5\footnotemark[1]&      6&     7\footnotemark[1]&     8\footnotemark[2]&     9\footnotemark[1]&    10&    11\footnotemark[1]&    12&    13 &  14\footnotemark[3]\\
       \hline
       $V_{n}\,({\mathrm{mV}})$& 6.0 & 7.8 & 10.5 & 11.2 & 13.3 & 14.5 & 16.0 & 17.1 & 19.6 & 20.6 & 21.3 & 23.0 & 25.7 & 29.1\\
       type&                      IR& Raman& IR& Raman&  Raman&    IR&   -&  -&    Raman&  IR&   -&  IR&  Raman &-\\
       symmetry&A$_\mathrm{2u}$&A$_\mathrm{1g}$&A$_\mathrm{2u}$&A$_\mathrm{1g}$&-&A$_\mathrm{2u}$&-&-&-&A$_\mathrm{2u}$&-&A$_\mathrm{2u}$&A$_\mathrm{1g}$&-\\

       assignment& Bi:CuCaSr & CuSr & SrCu & Sr:Cu & "disorder" & Ca:Sr & - & - & "disorder" & O3:O1 & - & O1:CaO3 & O1:Sr & -\\

       \end{tabular}
       \end{ruledtabular}
       \footnotetext[1]{very weak}
       \footnotetext[2]{disorder-induced IR-active or $B_{1g}$ Raman-active}
       \footnotetext[3]{disorder-induced IR-active or $A_{1g}$ (O3) Raman-active}
   \caption{\label{tab:phonons}
   Experimental voltage positions of phonon resonances. The mean values for all mesas at a Bi(Pb)-2212 crystal P22 are given. Also given are
types (Raman-active resp. IR-active), symmetries and assignments
of phonons.}
\end{table*}

Table \ref{tab:phonons} summarizes all observed phonon resonances for the
Bi(Pb)-2212 crystal. In total fourteen phonon resonances with
varying amplitudes could be distinguished. This is consistent with
the expected number of $c$-axis optical phonons in the tetragonal
Bi-2212 (seven IR and seven Raman) \cite{Kovaleva}. Indeed, by
comparison with optical spectroscopy \cite{Kovaleva}, most of the
observed phonons are easily identified with the known Raman and
IR-active $c$-axis optical phonons
\cite{Schlen,Ponomarev,Polariton}. Few unidentified phonons may be
disorder induced, due to the orthorhombic distortion, or the
incommensurate superstructure in the crystal \cite{Kovaleva}.

\begin{figure*}
\begin{center}
   \includegraphics[width=0.8\textwidth]{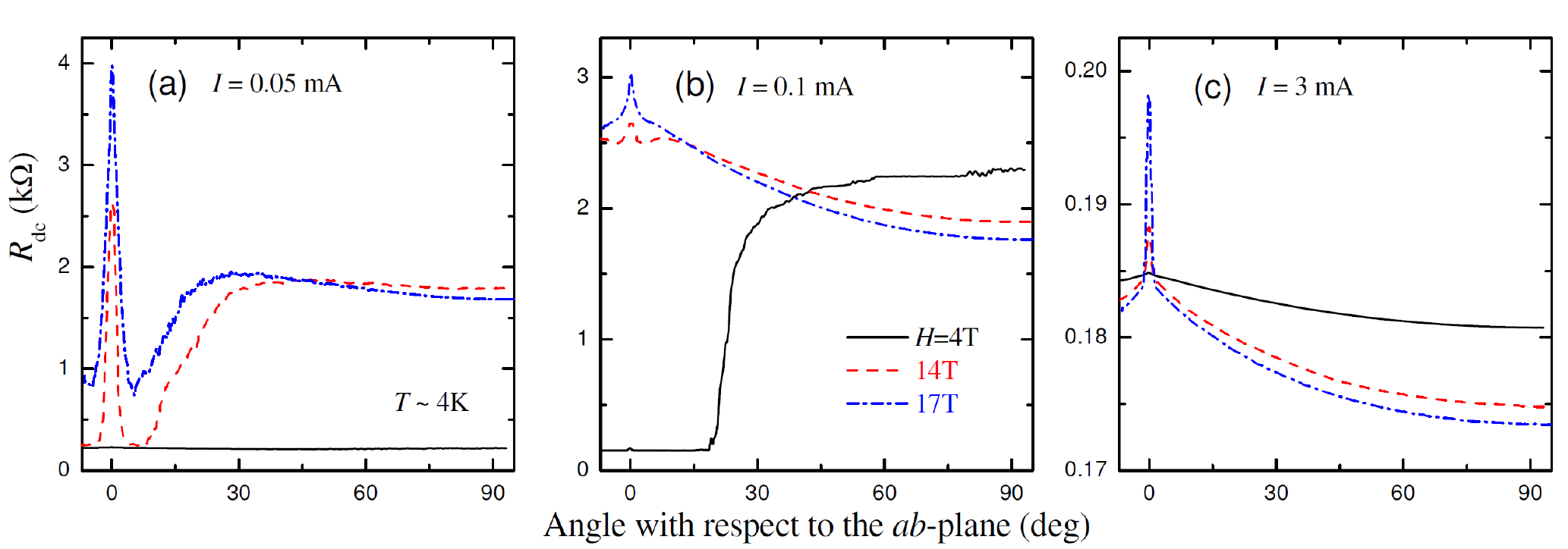}
\end{center}
\caption{\label{fig3} (Color online). Angular dependence of
dc-resistances, measured at different probe currents $I$ (a) low
$I=0.05\,\mathrm{mA}$, (b) moderate $I=0.1\,\mathrm{mA}$ and (c) high $I=3\,\mathrm{mA}$ for the
Bi(Pb)-2212 mesa P22/m3 at $T\simeq 4\,\mathrm{K}$ and at three different
magnetic fields 4\,T (solid), 14\,T (dashed) and 17\,T (dashed-dotted
line).}
\end{figure*}

From Fig.~\ref{fig1} it is obvious that an accurate sample alignment is
essential for the study of phonon resonances in intense fields.
Therefore we want to describe the developed unambiguous alignment
procedure in more details. Figure \ref{fig3} shows the angular dependence of the
dc-resistances $R_\mathrm{dc}=V/I$, measured at different probe currents
$I$ (a) low $I=0.05\,\mathrm{mA}$, (b) moderate $I=0.1\,\mathrm{mA}$ and (c) high
$I=3\,\mathrm{mA}$ for the Bi(Pb)-2212 mesa P22/m3 at $T\simeq 4\,\mathrm{K}$ and at
three different magnetic fields. The angular dependence of the
magnetoresistance can be understood from $I$-$V$s in Fig.~\ref{fig1}. The
sharp maximum at $\alpha =0$ is caused by the flux-flow
contribution. The flux-flow voltage is rapidly suppressed at small
a misalignment $\alpha \sim 1^{\circ}$ due to the pinning by
Abrikosov vortices. At larger angles the magnetoresistance
strongly depends on the bias: at a low bias the resistance increases
due to the suppression of the critical current, but at a higher bias
it is monotonously decreasing due to the negative QP
magnetoresistance \cite{MR,Bulaevskii}.

Even though the sharp flux-flow peak at low bias seems to be
advantageous for the sample alignment, we found that such an
alignment is unreliable because of the field lock-in along the
$ab$-plane, which may lead to a hysteresis up to a few degrees
\cite{LockIn}. In Fig.~\ref{fig3} (a) it can be seen as a slight asymmetry
of the minima in the $R_\mathrm{dc}(\alpha)$ curve at $H=17\,\mathrm{T}$. With
increasing field the angular range of the hysteresis decreases. At
$H=14\,\mathrm{T}$ it is only $\sim 0.2^{\circ}$. However this corresponds
to almost 500\,G $c$-axis field that is sufficient for penetration
of Abrikosov vortices. Therefore the low bias alignment is not
good enough. On the other hand, we observed that the alignment at
high bias, as shown in Fig.~\ref{fig3} (c), is working much better because
it is free from the hysteresis. This is caused by an effective
depinnig of Abrikosov (pancake) vortices as a combination of
self-heating that leads to thermal-assisted flux creep, a gradient
force due to perpendicular magnetization in pancake vortices with
respect to applied field and a drag force from Josephson vortices.
Furthermore, the maxima $R_\mathrm{dc}(\alpha)$ become significantly
sharper at high bias. For the curves at $H=14\,\mathrm{T}$ the full width at
half maximum is $3.4^{\circ}$ at low bias, Fig.~\ref{fig3} (a), compared to
only $1.5^{\circ}$ at high bias, Fig.~\ref{fig3} (c). Therefore we use high
bias magnetoresistance for the accurate sample alignment.

\begin{figure*}
\begin{center}
   \includegraphics[width=0.8\textwidth]{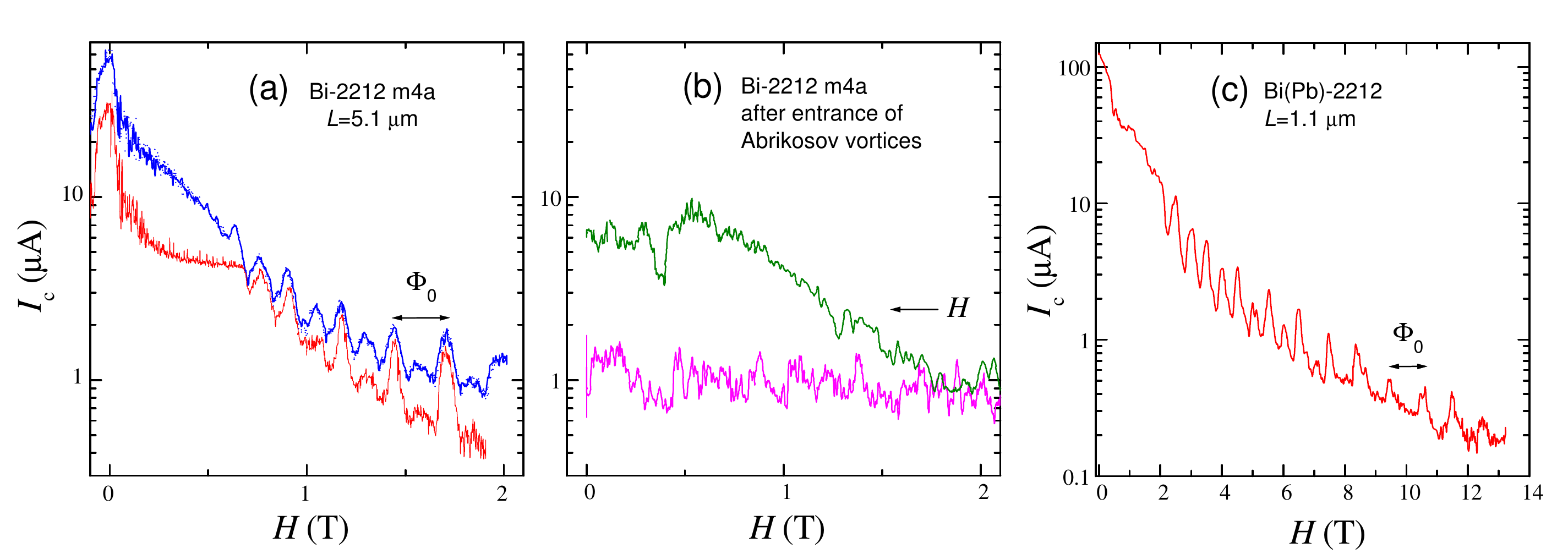}
\end{center}
\caption{\label{fig4} (Color online). Fraunhofer patterns $I_\mathrm{c}$ vs.
in-plane $H$ for a moderate size Bi-2212 mesa with the length
$L=5.1\,\mathrm{\mu m}$ at $T=1.6\,mathrm{K}$: (a) For field sweeps from zero to 2.1\,T
and backwards, (b) After trapping Abrikosov vortices at higher
field due to slight misalignment. Different curves in (a,b)
represent separate runs. (c) For a small Bi(Pb)-2212 mesa $L\simeq
1.2\,\mathrm{\mu m}$ after rigorous alignment at large bias. The clear
Fraunhofer modulation indicates the lack of Abrikosov vortex
intrusion up to 13\,T. Arrows indicate the field for introduction
of one flux quantum $\Phi_0$ in a single junction. A transition
from $\Phi_0/2$ modulation at low fields to $\Phi_0$ at large
fields is due to the reconstruction of the Josephson fluxon
lattice from triangular to rectangular.}
\end{figure*}

Figure \ref{fig4} shows the measured critical current vs the in-plane magnetic
field for (a,b) a moderate size Bi-2212 mesa and (c) a small
Bi(Pb)-2212 mesa P22/m4a. A clear Fraunhofer modulation is seen in
Figs.~\ref{fig4} (a) and (c). A transition from the half flux quantum to
the flux quantum $\Phi_0$ periodicity occurs with increasing field
due to the reconstruction of the Josephson fluxon lattice from the
triangular to the rectangular \cite{Katterwe,Koshelev}. Panel (b)
shows $I_\mathrm{c}(H)$ for the same mesa and alignment as in (a) after
subjecting the sample to a larger field $\sim 8\,\mathrm{T}$. Apparently, the
Fraunhofer modulation is completely destroyed by the intrusion of
Abrikosov vortices. In Fig.~\ref{fig4} (c) the Fraunhofer modulation
persists up to 13\,T, demonstrating that with our rigorous
alignment procedure we can prevent Abrikosov vortex intrusion even
at very high fields.

\begin{figure}
\begin{center}
   \includegraphics[width=0.4\textwidth]{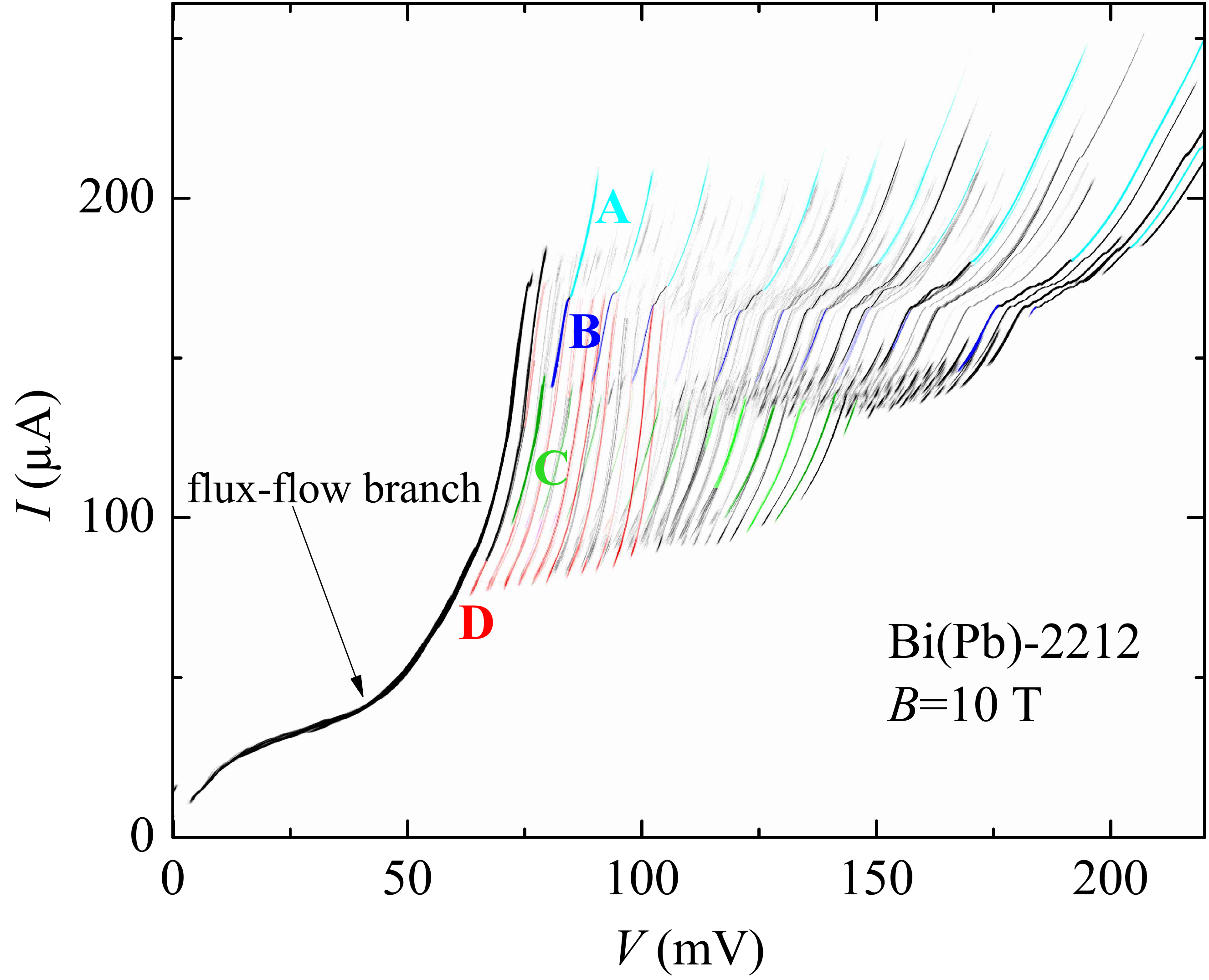}
\end{center}
\caption{\label{fig5} (Color online) $I$-$V$ oscillogram for the
Bi(Pb)-2212 mesa P22/m5b at $H=10\,\mathrm{T}$ aligned along the $ab$-plane.
Different families of phonon-polariton resonances are painted in
different colors. They are due to one-by-one joining of junctions
to the resonant state. The extraordinary large amplitude and
hysteresis of the resonances indicates their high quality factors
and oscillation amplitudes.}
\end{figure}

Figure \ref{fig5} shows the $I$-$V$ oscillogram for the Bi(Pb)-2212 mesa
P22/m5b at $T=1.6\,\mathrm{K}$ in the in-plane magnetic field of 10\,T. Here
a flux-flow branch ending with a nearly vertical Eck resonance is
clearly seen. Above it a large variety of phonon resonances
appears. Those resonances are divided into groups of $N=11$
branches due to one-by-one switching of IJJs into the resonant
state. For clarity we painted different families of branches by
different colors.

\begin{figure*}
\begin{center}
   \includegraphics[width=0.8\textwidth]{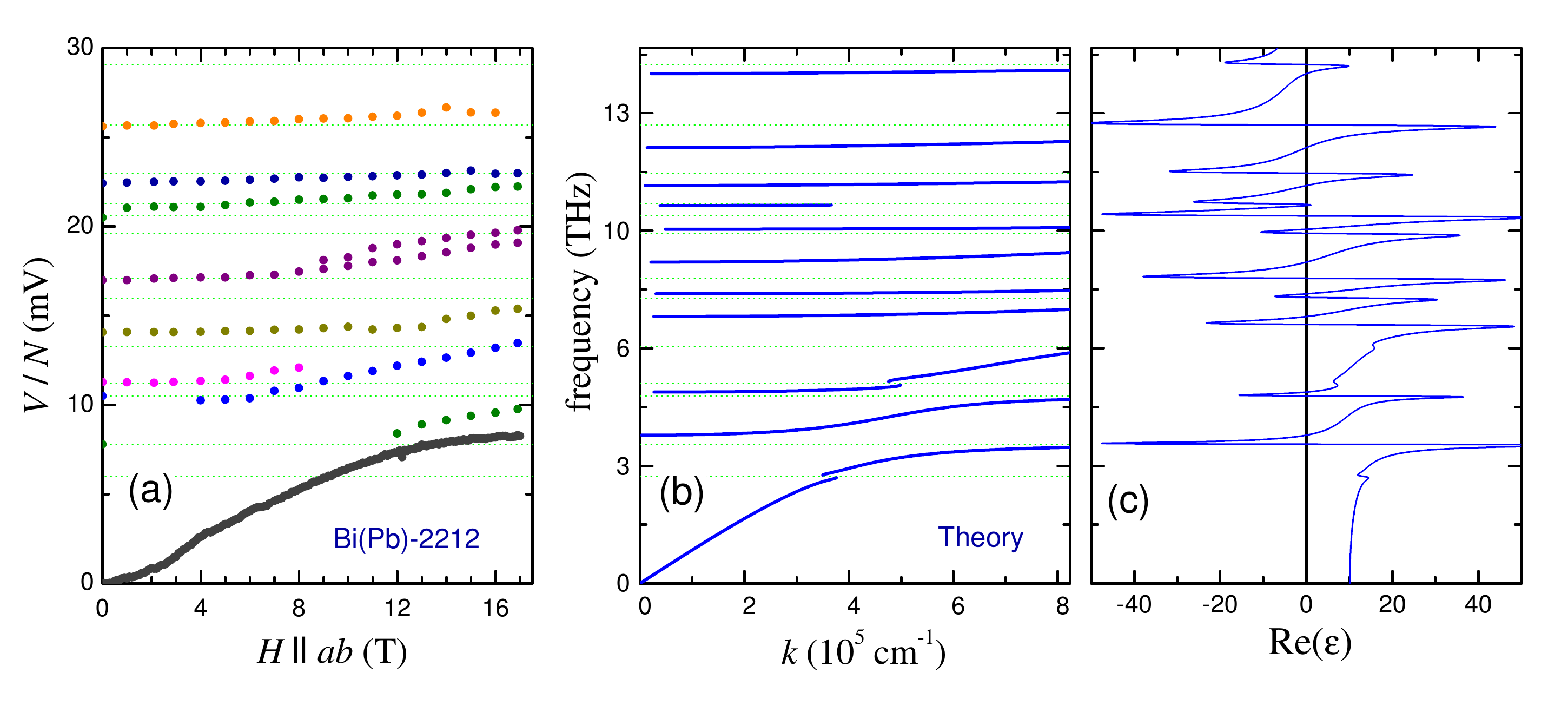}
\end{center}
\caption{\label{fig6} (Color online). (a) Magnetic field
dependence of voltages (per junction) of different resonances for
one of the Bi(Pb)-2212 mesas. It can be translated into the
dispersion relation $f(k)$ according to Eqs.~(\ref{Eq1}) and (\ref{Eq2}), which
correspond to the scales in panel (b). Positions of fourteen
characteristic phonon frequencies are marked by dotted lines.
Panels (b) and (c) represent the calculated dispersion relation
and the corresponding polaronic dielectric function, given by
Eq.~(\ref{eq:epsilon}).}
\end{figure*}

Figure \ref{fig6} (a) represents the magnetic field dependencies of
collective resonance voltages (all junctions at the same
resonance) for one of the studied Bi(Pb)-2212 mesas. Positions of
all fourteen phonon resonances are indicated by dotted lines. As
discussed in Ref.~\cite{Polariton}, the $V(H)$ plot can be
translated into the dispersion relation with the frequency given
by the ac-Josephson relation and the wave length by the
periodicity of the Josephson fluxon lattice:
\begin{eqnarray}\label{Eq1}
f=V/N\Phi_0,\\\label{Eq2}
k=2\pi H s/\Phi_0,
\end{eqnarray}
where $s=1.5\,\mathrm{nm}$ is the stacking periodicity of Bi-2212. Therefore
the $V$-$H$ plot represents the dispersion relation of
electromagnetic waves in the mesa. The lowest branch (black)
represents the Eck resonance. Initially it increases linearly with
field, corresponding to the constant speed of electromagnetic
waves $\simeq 4\times 10^5\,\mathrm{m/s}$, close to the out-of-phase speed
\cite{Superluminal}. However, at $H>12\,\mathrm{T}$ it saturates upon
approaching the second phonon mode. At this point the group
velocity of electromagnetic waves becomes very slow. This is a
well known signature for formation of phonon-polaritons with the
transverse optical phonon \cite{Preis}.

The polaronic dispersion relation is determined by the dielectric
function, which is influenced by the interaction of propagating
electromagnetic waves in the dielectric barrier with optical
phonons:

\begin{equation}
\varepsilon(\omega)=\varepsilon_\infty +
\sum_j{\frac{\omega_{\mathrm{TO},j}^2
S_j}{\omega_{\mathrm{TO},j}^{2}-\omega^{2}-\mathrm{i}\gamma_j \omega}}.
\label{eq:epsilon}
\end{equation}
Here $\omega_{\mathrm{TO},j}$ are frequencies of transverse optical
phonons, $S_j$ and $\gamma_j$ are the corresponding oscillator
strength and the damping parameter, respectively.

Figure \ref{fig6} (c) shows a simulated dielectric function with
characteristic phonon frequencies aligned to the positions of the
fourteen observed resonances. The oscillator strengths were chosen
so that $\varepsilon(0) =\varepsilon_{\infty}+\sum_j{S_j}$ is
equal to the experimental value $\sim 10$ and damping parameters
were chosen so that the corresponding dispersion relation $\omega
= c(0)k/\sqrt{\varepsilon(\omega)}$, shown in Fig.~\ref{fig6} (b) is
looking alike the experimental data in Fig.~\ref{fig6} (a). Figure \ref{fig6} (b) shows
the obtained dispersion relation for positive group velocities
with scales in frequency and the wave number corresponding to
voltage and field scales in Fig.~\ref{fig6} (a) according to Eqs.~(\ref{Eq1}) and
(\ref{Eq2}). Even though Fig.~\ref{fig6} (b) is not meant to be a direct fit to
Fig.~\ref{fig6} (a), it demonstrates that peculiarities of experimentally
observed $V(H)$ characteristics can be fairly well explained by
the simple expression Eq.~(\ref{eq:epsilon}). In particular, the saturation of
$V(H)$ and small group velocities are caused by the divergence of
the dielectric function when the ac-Josephson frequency is
approaching any of the active transverse-optical phonon
frequencies. Also an abrupt jump-like cancellation of resonances
at higher voltages (see Fig.~\ref{fig5}) is caused by the negative value of
$\varepsilon$, which prevents propagation of electromagnetic waves
with such frequencies and leads to switching of $I$-$V$s from the
resonant flux-flow to the non-resonant quasiparticle state.

In conclusion, we have described the accurate alignment procedure
and the angular dependence of phonon-polariton resonances between
electromagnetic waves, generated in the flux-flow state, and
$c$-axis optical phonons in Bi-2212 single crystals. The
phenomenon is caused by electrostriction in the insulating
SrO-2BiO-SrO layers. Our work has several implications. For
realization of the coherent flux-flow oscillator it is necessary
to stabilize the in-phase state with the rectangular Josephson fluxon
lattice, which is usually unstable because of fluxon-fluxon
repulsion. Strong interaction with infrared optical phonons can
stabilize the in-phase state. The stabilizing role of phonons is
seen already from the fact that we observed the flux-flow
phenomenon up to the record high frequencies $\sim 13\,\mathrm{THz}$, see
Fig.~\ref{fig6} (a) and (b). Formation of phonon-polaritons demonstrates
also that a residual polar response is left in cuprates even in
the superconducting state. This demonstrates that Bi-2212 single
crystals represent natural atomic metamaterials composed of
two-dimensional superconducting CuO$_2$-Ca-CuO$_2$ layers stacked
with polar insulating SrO-2BiO-SrO. This can lead to an extraordinary
strong electron-phonon coupling due to the combination of a long
range Coulomb interaction in 2D metallic planes with an unscreened
polar response in the adjacent polar insulator layers. This may
have significance for understanding the pairing mechanism in
cuprate superconductors.

Financial and technical support from the Swedish Research Council
and the SU-Core Facility in Nanotechnology is gratefully
acknowledged.

\end {document}